\documentclass[%
 reprint,
 amsmath,amssymb,
 aps,
prb,
]{revtex4-2}
\usepackage{graphicx}
\usepackage{dcolumn}
\usepackage{bm}
\usepackage{braket}
\usepackage{subcaption}
\usepackage{overpic}
\usepackage{xcolor}

\begin{document}

\title{Berry curvature quadrupole induced third-order anomalous Hall effect in altermagnetic $\mathrm{KV_2Se_2O}$}

\author{Jiacheng Yao}
\affiliation{Center for Quantum Matter, School of Physics, Zhejiang University, Hangzhou 310058, China.}
\author{Zhuocheng Lu}
\affiliation{Center for Quantum Matter, School of Physics, Zhejiang University, Hangzhou 310058, China.}
\author{Zhichao Guo}
\affiliation{Center for Quantum Matter, School of Physics, Zhejiang University, Hangzhou 310058, China.}
\author{Hua Wang}
\email{daodaohw@zju.edu.cn}
\affiliation{Center for Quantum Matter, School of Physics, Zhejiang University, Hangzhou 310058, China.}
\date{\today}

\begin{abstract}
Altermagnets combine compensated magnetic order with momentum-dependent spin splitting, providing a symmetry-rich platform for exploring nonlinear quantum-geometric transport. In the d-wave altermagnet considered here, the antiunitary symmetry operation $\mathcal{T}C_{4z}$ enforces a vanishing momentum-integrated Berry curvature, while inversion symmetry forbids a finite Berry-curvature dipole. Consequently, the Berry curvature quadrupole (BCQ) can provide a contribution to the nonlinear Hall response, scaling as $\tau^2$, where $\tau$ denotes the relaxation time.  We derive a gauge-covariant Wannier interpolation formula that enables efficient evaluation of the BCQ directly from first-principles Wannier tight-binding Hamiltonians. Applying this framework to $\mathrm{KV_2Se_2O}$, we calculate the momentum-space distribution of the BCQ and the associated chemical potential dependent nonlinear conductivity. The momentum-resolved analysis reveals that the dominant contribution originates from spin-orbit coupling induced avoided arossings, directly linking the gap structure to the sign-changing nonlinear response. Our work establishes a general first-principles route to higher-order quantum-geometric responses in altermagnets, opening avenues for exploring nonlinear Hall effects in next-generation spintronic and magnetoelectronic devices.
\end{abstract}

\maketitle


\section{\label{sec:introduction}Introduction}

Altermagnets constitute a symmetry-distinct class of collinear compensated magnets that combine vanishing net magnetization with momentum-dependent spin splitting \cite{PhysRevX.12.031042,PhysRevX.12.040501}. Unlike conventional antiferromagnets, their opposite-spin sublattices are connected by crystal rotations rather than solely by translations or inversion, allowing sizable nonrelativistic spin splitting even in the absence of spin-orbit coupling (SOC) \cite{PhysRevX.12.011028,PhysRevLett.126.127701}. This electronic structure is governed by spin-space symmetries \cite{PhysRevX.14.031037,Yang2024,PhysRevX.14.031039} and enables unconventional spin-current and spin-caloritronic responses \cite{Ma2021Multifunctional,PhysRevB.111.035423}. When SOC is included, the reduced magnetic symmetries can further permit crystal and anomalous Hall effects \cite{Smejkal2020,Feng2022}. These properties make altermagnets a promising setting for exploring how magnetic symmetry selects the lowest nonvanishing order of nonlinear transverse transport.

For Berry-curvature-driven Hall responses, this symmetry-based order selection can be organized through a hierarchy of momentum-space multipoles. The linear intrinsic anomalous Hall effect is
governed by the Brillouin-zone-integrated Berry curvature; when this
quantity vanishes by symmetry, its dipole moment may remain finite and
generate a second-order nonlinear Hall response
\cite{Xiao2010BerryPhase,PhysRevLett.115.216806,Ma2019Observation}.
If symmetry further eliminates the Berry-curvature dipole, the
Berry-curvature quadrupole (BCQ) can provide the leading
contribution to the third-order Hall conductivity, scaling as $\tau^2$
\cite{PhysRevB.107.115142}, where $\tau$ denotes the relaxation time. Experimentally, evidence for a BCQ-induced third-order nonlinear Hall response has been reported in the antiferromagnet FeSn, while quadrupolar quantum-geometric transport has also been observed in MnBi$_2$Te$_4$ \cite{Sankar2024BCQFeSn,Li2024QuantumGeometryQuadrupole}.
Recent symmetry and low-energy-model studies have shown that such third-order Hall responses are allowed in altermagnets, with model calculations suggesting that they can be induced by spin-orbit-coupling-driven anticrossing structures in the spin-split bands \cite{PhysRevLett.133.106701, Xiang2026ThirdOrderIAHEAltermagnets}.
Although realistic first-principles-derived tight-binding models have recently been developed for the V$_2$Se$_2$O family \cite{Cheng2026V2Se2OTightBinding}, a consistent evaluation of the BCQ from DFT-derived Wannier Hamiltonians requires an analytical formulation of the second momentum derivatives of the Berry curvature in the Wannier basis.

In this work, building on established Wannier-interpolation and covariant derivative frameworks \cite{PhysRevB.74.195118, liu2023covariantderivativesberrytypequantities}, we derive an explicit Wannier-interpolation formula for the BCQ tensor in multiband systems. The derivation expresses the second covariant momentum derivatives of the Berry curvature in terms of Wannier-gauge quantities, including the basis-transformation and interband-connection corrections required in the energy-eigenstate representation. This formulation enables efficient evaluation of the BCQ and its associated $\tau^2$ contribution to the third-order Hall conductivity on dense three-dimensional Brillouin-zone meshes. The implementation is benchmarked against direct finite-difference calculations and analytical results for a minimal altermagnetic model, and is subsequently applied to a first-principles-derived Wannier Hamiltonian.

As a representative material system, we consider KV\(_2\)Se\(_2\)O, in which room-temperature \(d\)-wave spin-momentum locking has been reported \cite{Jiang2025}. In our first-principles calculations, we adopt the C-type magnetic configuration, which supports the characteristic \(d\)-wave altermagnetic band structure \cite{Sun2025KV2Se2ONeutron,Gu2026KV2Se2OMagneticConfigurations}. Using the derived BCQ formulation, we resolve the band- and momentum-resolved contributions to the BCQ. Our results reveal distinct contributions from two characteristic features of the electronic structure: the extended altermagnetic nodal-line regions generate a broad BCQ background, whereas the dominant BCQ peak originates from an SOC-induced avoided crossing between a specific pair of bands.

\section{\label{sec:symmetry}Symmetry constraints on Berry-curvature multipoles}

We first clarify the symmetry setting relevant to the BCQ contribution considered in this work. 
The nonrelativistic altermagnetic band structure and the relativistic Berry-curvature response are governed by two related but distinct symmetry descriptions. 
In the absence of SOC, the C-type magnetic configuration of KV$_2$Se$_2$O is described by a spin group in which the two opposite-spin V sublattices are connected by the combined spin-space operation
$[\mathcal{K}\Vert C_{4z}]$,
where $\mathcal{K}$ reverses the collinear spin polarization and $C_{4z}$ rotates the crystal momentum by $90^\circ$ about the $z$ axis. 
The operation connects the two spin sectors through the relation 
$E_{\uparrow}(k_x,k_y,k_z)
=
E_{\downarrow}(-k_y,k_x,k_z)$,
and produces the characteristic $d$-wave momentum-dependent spin splitting. 
It simultaneously protects the nonrelativistic spin-degenerate nodal structures along the corresponding symmetry-invariant momentum lines, in agreement with the C-paired spin-momentum-locking
classification of KV$_2$Se$_2$O \cite{Hu2025CPairedSML}.
Such spin-group constraints characterize the altermagnetic band structure before SOC is introduced and are distinguished from the magnetic-group constraints on the relativistic Berry curvature discussed below.

For the C-type configuration with the N\'eel vector along the
crystallographic $c$ axis, the magnetic space group is
$P4^{\prime}/mm^{\prime}m$ \cite{Yan2026InverseMagneticBreakdown},
corresponding to the magnetic point group $4^{\prime}/mm^{\prime}m$.
The relevant symmetry operations include inversion $\mathcal P$ and the antiunitary fourfold rotation
$4_z^{\prime}\equiv\mathcal{T}C_{4z}$,
where $\mathcal T$ denotes time reversal. 
SOC therefore reduces the nonrelativistic spin group to a magnetic space group, but the remaining magnetic symmetries continue to impose constraints on the Brillouin-zone moments of the Berry curvature. For the transverse Hall channel considered here, we denote the relevant Berry-curvature component by
$\Omega^{xy}_{n}(\mathbf{k})\equiv\Omega_{n,z}(\mathbf{k})$.
Under the antiunitary operation $4_z^{\prime}$, this component transforms as
\begin{equation}
\Omega^{xy}_{n}(k_x,k_y,k_z)
=
-\Omega^{xy}_{\bar n}(k_y,-k_x,-k_z),
\label{eq:bc_c4t}
\end{equation}
where $\bar n$ denotes the symmetry-related band. 
Equation~(\ref{eq:bc_c4t}) does not require the Berry curvature to vanish locally; rather, it connects momentum-space regions carrying opposite Berry curvature, so that their contributions cancel and the summed Berry curvature vanishes.

To identify the allowed integrated Berry-curvature moments, we define
\begin{align}
B^{xy}
&=
\sum_n\int_{\mathrm{BZ}}[d\mathbf{k}]\,
f_{0n}\Omega^{xy}_{n}(\mathbf{k}),
\label{eq:bc_monopole}
\\
D^{xy}_{\alpha}
&=
\sum_n\int_{\mathrm{BZ}}[d\mathbf{k}]\,
f_{0n}\partial_{\alpha}\Omega^{xy}_{n}(\mathbf{k}),
\label{eq:bcd}
\\
Q^{xy}_{\alpha\beta}
&=
\sum_n\int_{\mathrm{BZ}}[d\mathbf{k}]\,
f_{0n}\partial_{\alpha}\partial_{\beta}
\Omega^{xy}_{n}(\mathbf{k}),
\label{eq:bcq}
\end{align}
where $[d\mathbf{k}]\equiv d^3k/(2\pi)^3$, 
$f_{0n}$ is the equilibrium occupation, and
$\partial_{\alpha}\equiv\partial/\partial k_{\alpha}$. The antiunitary fourfold symmetry maps the integrand of
Eq.~(\ref{eq:bc_monopole}) onto a symmetry-related contribution with the opposite sign. 
Consequently, $B^{xy}=0$,
and the corresponding intrinsic linear anomalous Hall contribution is forbidden. 
Inversion symmetry independently eliminates the Berry-curvature dipole. Owing to the axial nature of the Berry curvature,  $\Omega^{xy}$ is even under inversion,
$\Omega^{xy}_{n}(\mathbf{k})
=
\Omega^{xy}_{\tilde n}(-\mathbf{k})$,
where $\tilde n$ is the inversion-related band. 
The first momentum derivative of $\Omega^{xy}$ is therefore odd,
$\partial_{\alpha}\Omega^{xy}_{n}(\mathbf{k})
=
-\partial_{\alpha}\Omega^{xy}_{\tilde n}(-\mathbf{k})$,
yielding
$D^{xy}_{\alpha}=0$.
Hence the Berry-curvature-dipole contribution to the second-order nonlinear Hall response is forbidden even with SOC in altermagnetic KV$_2$Se$_2$O. 
This symmetry constraint is consistent with recent analyses showing that the
combined $C_4\mathcal T$ symmetry forbids both the linear Hall response and the
equilibrium Berry-curvature-dipole response in a pure $d$-wave altermagnet
\cite{Mukherjee2026SecondOrderAHE}.

By contrast, the second momentum derivative of the Berry curvature is even under inversion and is not forced to vanish. 
The antiunitary fourfold rotation constrains the in-plane components according to
$
Q^{xy}_{xx}=-Q^{xy}_{yy}$,
or, in the Cartesian-index notation used in the numerical output,
$Q_{1211}=-Q_{1222}$. 
Here, the antisymmetric upper index pair of Q originates from the Berry curvature and therefore transforms as an axial vector, whereas the lower indices, arising from momentum derivatives, transform as polar-vector components. The mirror symmetries of $4^{\prime}/mm^{\prime}m$ eliminate the off-diagonal in-plane components, requiring $Q^{xy}_{xy}=Q^{xy}_{yx}=0$. The surviving combination $Q^{xy}_{d}\equiv Q^{xy}_{xx}=-Q^{xy}_{yy}$ captures the leading nonlinear response. 
In the absence of SOC, time-reversal symmetry is already broken by the magnetic order, so the Berry curvature can be locally nonzero; however, the nonrelativistic spin-group symmetries enforce that all Brillouin-zone moments vanish identically, eliminating any transverse Berry-curvature response. Once SOC is switched on, the spin-group structure is reduced to the magnetic space group and a nontrivial local Berry curvature develops. The antiunitary $4_z^{\prime}$ symmetry still enforces cancellation of its Brillouin-zone integral, while inversion $\mathcal{P}$ (together with the resulting $C_2$ symmetry) eliminates the dipole moment. The quadrupole therefore emerges as the lowest surviving multipole contribution to transverse transport.

\section{\label{sec:formalism}Third-order transport formalism}

 The purpose of this section is to present the nonlinear transport formula and the Wannier-interpolation framework; implementation details and numerical benchmarks are provided in the Supplemental Material~\cite{SM}. We focus exclusively on the BCQ contribution to the transverse response and do not evaluate the quantum-metric-quadrupole sector, which is more naturally associated with longitudinal components in the altermagnetic context discussed in Ref.~\cite{PhysRevLett.133.106701}.

We start from the semiclassical wave-packet dynamics in a uniform electric field. Denoting the wave-packet position and crystal momentum by $\mathbf{r}$ and $\mathbf{k}$, respectively, the velocity acquires an anomalous contribution from the Berry curvature~\cite{Xiao2010BerryPhase},
\begin{align}
\dot{\mathbf{r}}=\mathbf{v}-\dot{\mathbf{k}}\times\bm{\Omega},
\qquad
\hbar\dot{\mathbf{k}}=-e\mathbf{E},
\end{align}
where $\mathbf{v}$ is the band velocity and $\bm{\Omega}$ is the Berry curvature.

Within the relaxation-time approximation, the distribution function $f$ satisfies
\begin{align}\label{eq:Eq.1}
    \frac{\partial{f}}{\partial t}+\dot{\mathbf{k}}\cdot \frac{\partial f}{\partial \mathbf{k}}=-\frac{f-f_{0}}{\tau},
\end{align}

where $f_{0}$ is the equilibrium Fermi-Dirac distribution function and $\tau$ is taken to be a constant. Treating the applied electric field as a perturbation, we expand the distribution function as
\begin{align}
    f=f_{0}+f_{1}+f_{2}+f_{3}.
\end{align}
The electric current density is given by
\begin{equation}
j_\alpha=-e\int_{\mathbf{k}}\dot r_\alpha f(\mathbf{k}),
\end{equation}
where $\int_{\mathbf{k}}\equiv \int d^3k/(2\pi)^3$. Expanding the current to third order in the electric field, one obtains a conventional Fermi-surface contribution and a Berry-curvature-induced contribution. The latter originates from the anomalous velocity and can be characterized by the BCQ tensor,
\begin{equation}
Q^{\alpha\beta}_{\gamma\delta}
=
\sum_n
\int_{\mathbf k}
f_{0n}\,
\partial_\gamma\partial_\delta
\Omega_n^{\alpha\beta}(\mathbf{k}) .
\label{eq:BCQ_sea}
\end{equation}

Using integration by parts, Eq.~(\ref{eq:BCQ_sea}) can be rewritten in the equivalent Fermi-surface form
\begin{equation}
Q^{\alpha\beta}_{\gamma\delta}
=
\hbar\sum_n\int_{\mathbf k}
\partial_\gamma
\Omega_n^{\alpha\beta}(\mathbf{k})
\,v_{n}^{\delta}\,
\delta(E_n-\mu),
\label{eq:BCQ_surface}
\end{equation}
where $\Omega_n^{\alpha\beta}$ is the Berry curvature of band $n$,
$v_{n}^{\delta}=\hbar^{-1}\partial_{k_\delta}E_n$ is the
physical group velocity, $E_n$ is the band energy, and $\mu$ is the chemical
potential.  Our implementation instead returns the energy gradient
$v_n^\delta=\partial_{k_\delta}E_n$; with that convention the explicit factor
of $\hbar$ in Eq.~(\ref{eq:BCQ_surface}) is absent.  Equation~(\ref{eq:BCQ_surface})
shows that the BCQ is governed by the momentum-space variation of Berry
curvature on the Fermi surface.

With the convention $j_a^{(3)}=\sigma_{abcd}^{(3)}E_bE_cE_d$, the BCQ part of the third-order conductivity follows the $\tau^2$ .  The explicitly field-symmetrized tensor is
\begin{align}
\sigma^{(3),\mathrm{BCQ}}_{abcd}
={}&-\frac{e^4\tau^2}{3\hbar^3}
\sum_n\int_{\mathbf{k}} f_0
\Bigl(
\partial_c\partial_d\Omega_n^{ab}
\notag\\
&\hspace{30mm}
+\partial_b\partial_d\Omega_n^{ac}
+\partial_b\partial_c\Omega_n^{ad}
\Bigr),
\label{eq:BCQ_conductivity}
\end{align}
where the three terms enforce permutation symmetry among $b,c,d$.  After contraction with three identical static fields, the three equivalent terms related by dummy index relabeling collapse to a single expression with no additional factor of $1/3$. The third-order conductivity generally comprises scattering time independent intrinsic terms, $\tau$-linear quantum metric quadrupole contributions, and the conventional $\tau^3$ Drude response \cite{PhysRevB.107.075411,10.21468/SciPostPhys.14.4.082}; we do not evaluate these additional contributions in the present work.

\subsection{\label{subsec:berry}Berry curvature and covariant derivatives}

For the Bloch eigenstate $|u_n(\mathbf{k})\rangle$ of band $n$, the Berry connection is defined as
\begin{equation}
A_n^\alpha(\mathbf{k})
=
i\langle u_n(\mathbf{k})
|
\partial_\alpha
u_n(\mathbf{k})
\rangle ,
\end{equation}
where $\partial_\alpha\equiv\partial/\partial k_\alpha$. The Berry curvature is then given by
\begin{equation}
\Omega_n^{\alpha\beta}(\mathbf{k})
=
\partial_\alpha A_n^\beta(\mathbf{k})
-
\partial_\beta A_n^\alpha(\mathbf{k}).
\label{eq:berry_curvature}
\end{equation}

As a gauge-invariant quantity, the Berry curvature encodes the geometric structure of Bloch states in momentum space. In multiband systems, ordinary derivatives generally contain gauge-dependent contributions within the chosen subspace, particularly in the vicinity of band degeneracies. Capturing physically meaningful geometric responses therefore requires replacing the ordinary derivative with its covariant counterpart \cite{liu2023covariantderivativesberrytypequantities}. The covariant derivative of an operator $X$
projected onto a chosen degenerate band subspace $\mathcal{S}$ is defined as
\begin{equation}
X^{;\alpha}_{nm}
=
X^{,\alpha}_{nm}
-
D^{\alpha}_{nl}X_{lm}
+
X_{nl}D^{\alpha}_{lm},
\qquad
n,m\in\mathcal{S},\quad l\in\bar{\mathcal{S}},
\label{eq:covariant}
\end{equation}
where summation over the repeated index $l$ is implied and
\begin{equation}
X^{,\alpha}_{nm}
=
\langle n|\partial_\alpha X|m\rangle .
\end{equation}
Here, $\bar{\mathcal{S}}$ denotes the complementary subspace to
$\mathcal{S}$. The interband connection between the two subspaces is given by
\begin{equation}
D^{\alpha}_{ln}
=
\frac{\langle l|\partial_\alpha H|n\rangle}
{E_n-E_l},
\qquad
n\in\mathcal{S},\quad
l\in\bar{\mathcal{S}},
\label{eq:Dmn}
\end{equation}
which is the off-diagonal matrix element of the momentum derivative of $H$ in the eigenstate basis, i.e., the first-order interband mixing amplitude induced by $\partial_\alpha$. It enters the covariant derivative of operators defined in $\mathcal{S}$ as the correction arising from the frozen outer subspace $\bar{\mathcal{S}}$.

Throughout, all momentum derivatives of the Berry curvature are evaluated within the covariant-derivative formalism. Notation follows this distinction: a comma denotes an ordinary momentum derivative evaluated in the Wannier gauge, whereas a semicolon denotes the corresponding covariant derivative in the energy-eigenstate basis, so that $\Omega^{\alpha\beta,\gamma\delta}$ and $\Omega^{\alpha\beta;\gamma\delta}$ represent the two operations.

\subsection{\label{subsec:wannier}Wannier representation of the BCQ}

To evaluate the BCQ efficiently over dense Brillouin-zone meshes, we rewrite the geometric quantities in a Wannier-interpolation framework. The central task is to express momentum derivatives in the energy eigenstate basis while keeping track of the Berry connection inherited from the Wannier basis. For the multiband transformation, the derivative of a Bloch eigenstate can be written as
\begin{align}
\partial_\alpha\ket{u_n}
=
\ket{\partial_\alpha u_j}U_{jn}
+
\ket{u_j}\partial_\alpha U_{jn},
\end{align}
where $U_{jn}$ denotes the unitary transformation from the reference basis, indexed by $j$, to the energy eigenstate basis, indexed by $n$, with $j$ running over the full reference-basis Hilbert space.

\begin{align}
\braket{u_n|\partial_\alpha u_{l}}
&=
U_{nj}^{\dagger}
\braket{u_{j}|\partial_\alpha u_j}
U_{jl}
+
U_{nj}^{\dagger}\partial_\alpha U_{jl}
\nonumber 
\\&=
\frac{
\bra{n}\partial_\alpha H\ket{l}
}{
E_{l}-E_n
}
\\&=-iA^{J,\alpha}_{nl}+\frac{
H^{J,\alpha }_{nl}
}{
E_{l}-E_n
},\qquad n\in\mathcal{S},\quad
l\in\bar{\mathcal{S}},
\end{align}

Here $A^J$ and $H^J$ denote quantities first evaluated in the
reference (Wannier) basis and then transformed to the energy eigenstate
basis by the unitary matrix $U$. For an operator $O$, we use the notation
\begin{equation}
O^J_{\rm eig}=U^\dagger O^J U ,
\end{equation}
where $J$ labels the reference-basis representation and ``eig'' denotes the
energy eigenstate representation. In particular,
\begin{equation}
H^{J,\alpha }_{nl}
=
U^\dagger
\partial_\alpha
\langle u_{j'}|H|u_j\rangle
U .
\end{equation}
The corresponding interband connection is
\begin{equation}
D^{J,\alpha}_{nl}
\equiv
\frac{
H^{J,\alpha }_{nl}
}{
E_{l}-E_n
},
\qquad n\in\mathcal{S},\quad
l\in\bar{\mathcal{S}} .
\end{equation}
For an operator $O$ expressed in the energy eigenstate basis, the 
derivative induced by the reference-basis interpolation is
\begin{align}
\partial_{\alpha}(U^{\dagger}_{nj}O^{J}_{jj'}U_{j'n})\label{eq:covar2}
=
O^{J,\alpha}_{\rm eig}
-
D^{J,\alpha}O^J_{\rm eig}
+
O^J_{\rm eig}D^{J,\alpha}.
\end{align}
By defining
\begin{align}
D_{nm}^{J,\alpha}
=
iA_{nm}^{J,\alpha},
\qquad n,m\in\mathcal{S},
\end{align}
the gauge-dependent terms generated by the derivative
$\partial_{\alpha}O^{J}$ within the degenerate subspace $\mathcal{S}$ are exactly canceled by the corresponding connection
terms in Eq.~\ref{eq:covar2}. Explicitly,
\begin{align}
O^{J,\alpha}_{{\rm eig},nm}
=
iA^{J,\alpha}_{nj}O_{jm}
+
O^{,\alpha}_{nm}
-
iO_{nj}A^{J,\alpha}_{jm}.
\end{align}
Equation~\eqref{eq:covar2} then reduces to the covariant-derivative form of Eq.~\eqref{eq:covariant}, which we denote by $O^{J;\alpha}_{\rm eig}$. The calculation thus reduces to evaluating the ordinary momentum derivative in the $J$ representation together with the momentum derivatives of the transformation matrix $U$.

For practical Wannier interpolation, $D^{\alpha}$ must be re-expressed in terms of Wannier-gauge quantities, from which an explicit interpolation formula can be derived. The full derivation is provided in the Supplemental Material~\cite{SM}. All BCQ results reported below are obtained from the resulting compact covariant-derivative expression, with finite-difference calculations used solely as numerical benchmarks.The explicit Wannier-interpolation expression is given below.
All matrices are represented in the energy-eigenstate basis,
while comma derivatives are evaluated in the Wannier gauge and
subsequently transformed to this basis. The $W$ labels associated
with this intermediate Wannier-gauge evaluation are omitted for brevity.
\begin{align}
\Omega^{\alpha\beta;\gamma\delta}
={}&\Omega^{\alpha\beta,\gamma\delta}
+\left[\Omega^{\alpha\beta,\delta},D^\gamma\right]
+\left[\Omega^{\alpha\beta,\gamma},D^\delta\right]
\notag\\
&-D^\delta\Omega^{\alpha\beta}D^\gamma
-D^\gamma\Omega^{\alpha\beta}D^\delta
\notag\\
&+\bigl(D^{\gamma,\delta}\bigr)^\dagger
  \Omega^{\alpha\beta}
+\Omega^{\alpha\beta}D^{\gamma,\delta}
\notag\\
&-\Bigl(
 A^\alpha D^\beta
 +(D^\beta)^\dagger A^\alpha
 -i(D^\alpha)^\dagger D^\beta
 \Bigr)^{,\gamma\delta}
\notag\\
&-\Bigl(
 (D^\gamma)^\dagger A^\alpha D^\beta
 +(D^\beta)^\dagger A^\alpha D^\gamma
 \Bigr)^{,\delta}
\notag\\
&+D^\delta\Bigl(A^\alpha D^\beta\Bigr)^{,\gamma}
-\Bigl((D^\beta)^\dagger A^\alpha\Bigr)^{,\gamma}D^\delta
\notag\\
&-(\alpha\leftrightarrow\beta).
\label{eq:bcq-compact}
\end{align}
For example, the second-order momentum derivative of the Berry curvature $\Omega^{\alpha\beta,\gamma\delta}$ is evaluated as
\begin{align}
\Omega^{\alpha\beta,\gamma\delta} 
= U^{\dagger}\left[\partial_{\gamma}\partial_{\delta}\left(
 \partial_{\alpha} A^{\beta} - \partial_{\beta} A^{\alpha} \right)\right]^{W}U,
\end{align}
where the superscript $W$ indicates evaluation in the Wannier gauge,
and the factors $U^{\dagger}$ and $U$ transform the resulting matrix
to the energy-eigenstate basis.

Unless otherwise specified,matrix multiplication implies summation over the full Hilbert space,with $n\in S$ and $l\in\bar S$ denoting states inside and outside the selected subspace, respectively. Contraction indices in matrix products are also suppressed.The notation $(\alpha\leftrightarrow\beta)$ denotes the interchange of $\alpha$ and
$\beta$ in the indicated terms, except in $\Omega^{\alpha\beta}$, which is
kept unchanged. Equation~\eqref{eq:bcq-compact}  can also be rewritten entirely in terms of Wannier-gauge quantities. In particular,
\begin{align}
\left(\partial_{\alpha}A^{\beta}-\partial_{\beta}A^{\alpha}\right)^W
=
\braket{\partial_{\alpha}u^W|\partial_{\beta}u^W}
-
\braket{\partial_{\beta}u^W|\partial_{\alpha}u^W},
\end{align}
which corresponds to the Wannier-gauge combination
$F^W_{\alpha\beta}-F^W_{\beta\alpha}$ introduced in
Ref.~\cite{liu2023covariantderivativesberrytypequantities}. That work establishes a general framework for expressing higher-order covariant derivatives in terms of Wannier-gauge quantities and for constructing a broad class of geometric responses. For the BCQ here, the nested structure of the covariant derivatives generates numerous terms when expanded to second order. To enable a more direct and efficient implementation, we instead construct the required derivatives explicitly from the Wannier-gauge quantities $A^W$ and $H^W$ supplied by the Wannier tight-binding model.

More explicitly, the derivative of $D^{\beta}$ can be expressed as
\begin{align}
\partial_{\alpha}D^{\beta}
=
U^{\dagger}\partial_{\alpha}\partial_{\beta}U
-
D^{\alpha}D^{\beta}.
\label{eq:diag}
\end{align}
In our derivation, the basis transformation $U$ and its momentum derivatives serve as the fundamental objects. This formulation allows higher-order connection terms to be expressed directly in terms of derivatives of $U$, thereby avoiding
the proliferation of repeated products of interband connections. For instance,
a product such as
\begin{align}
D^{\alpha}D^{\beta}
\sim
U^{\dagger}_{nj}(\partial_{\alpha}U)_{jm}
\,U^{\dagger}_{mj'}(\partial_{\beta}U)_{j'n'}
\end{align}
can be incorporated into a more compact representation. We therefore introduce
the higher-order quantity
\begin{align}
D^{\alpha_{1},\alpha_{2},\ldots,\alpha_{n}}
&\equiv
U^{\dagger}
\left(
\partial_{\alpha_{1}}
\partial_{\alpha_{2}}
\cdots
\partial_{\alpha_{n}}U
\right),
\\
\left(
D^{\alpha_{1},\alpha_{2},\ldots,\alpha_{n}}
\right)^{\dagger}
&=
\left(
\partial_{\alpha_{1}}
\partial_{\alpha_{2}}
\cdots
\partial_{\alpha_{n}}U^{\dagger}
\right)U,
\end{align}
which packages the higher-order momentum variation of the basis transformation into a single tensor and markedly reduces the number of explicit multi-$D$ terms in the final expression. The quantities $D^{\alpha_{1},\alpha_{2},\ldots,\alpha_{n}}$ can thus be treated as independent building blocks, without expanding them into products of lower-order connection matrices. The resulting formulation enables efficient evaluation of the BCQ on dense Brillouin-zone meshes while bypassing finite differences of gauge-dependent band quantities. Direct finite-difference calculations are retained only as an independent numerical benchmark.

To further validate the formalism, we consider a minimal altermagnetic model Hamiltonian parameterized by the altermagnetic exchange and the SOC strength. Applying the same numerical framework to this model yields momentum-resolved BCQ distributions in excellent agreement with the corresponding analytical results, providing a controlled benchmark of both the derived formulation and the numerical implementation~\cite{SM}. As reported in the timing tests in the Supplemental Material, the Wannier-interpolation implementation achieves roughly a $2.2\times$ speedup over the finite-difference scheme on a small $200\times200$ mesh.

\section{\label{sec:computational}Computational details}

First-principles calculations for bulk $\mathrm{KV_2Se_2O}$ were performed using the Vienna \textit{ab initio} Simulation Package (VASP) \cite{Kresse1996EfficientIterativeSchemes} within the projector augmented-wave method \cite{Blochl1994ProjectorAugmentedWave,Kresse1999PAWUltrasoft}. We adopted the experimental tetragonal structure (space group $P4/mmm$) reported at $T=293~\mathrm{K}$, with lattice parameters $a=b=3.946~\text{\AA}$ and $c=7.312~\text{\AA}$ \cite{Yan2026SpinValley}. The exchange-correlation functional was treated within the Perdew--Burke--Ernzerhof generalized-gradient approximation \cite{Perdew1996PBE}. A plane-wave cutoff energy of $600$ eV and a $\Gamma$-centered $6\times6\times9$ $k$-point mesh were employed, with spin-orbit coupling (SOC) included in the calculations.

 Maximally localized Wannier functions constructed from V $d$, O $p$, and Se $p$ orbitals were generated using Wannier90 \cite{Marzari1997MaximallyLocalizedWannier,Mostofi2008Wannier90}. The resulting symmetry-corrected Wannier Hamiltonian with $U_{\rm eff}=0$ eV was used for the BCQ calculations, with implementation and validation details provided in the Supplemental Material.


The two-dimensional momentum-resolved BCQ distributions were evaluated on a $400\times400$ interpolation mesh, while the Brillouin-zone integration of the nonlinear conductivity was carried out on a dense $2000\times2000\times100$ $k$-point mesh. Both calculations were performed using WannierNLQG \cite{Lu2026WannierNLQG}, an in-house first-principles post-processing code developed and openly released by the authors, in which the Wannier-interpolation formulation of BCQ derived in this work has been implemented. Convergence tests are documented in the Supplemental Material~\cite{SM}.

\section{\label{sec:material_results}Material platform and microscopic mechanism}

\begin{figure}[t]
\centering
\includegraphics[width=\linewidth]{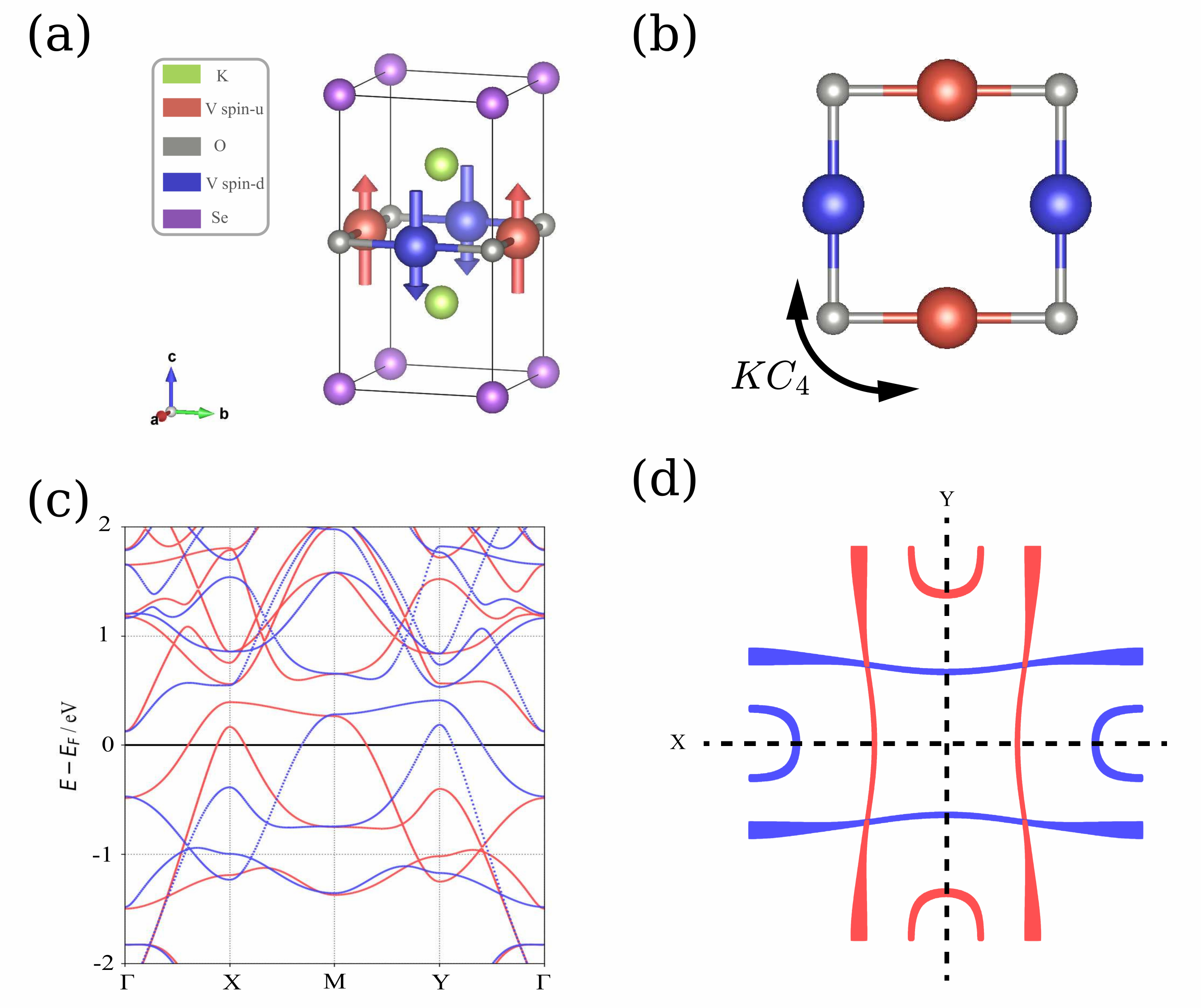}

\caption{\label{fig:structure_bands}Crystal, magnetic, and electronic structure of $\mathrm{KV_2Se_2O}$. (a) Crystal and magnetic structure. Red and blue arrows indicate opposite V spin orientations. (b) Top view of the $\mathrm{V_2O}$ plane, where opposite-spin V sublattices are related by the altermagnetic spin-space operation. (c) Spin-resolved band structure without SOC along high-symmetry lines. Red, blue, and green curves denote spin-up, spin-down, and spin-degenerate bands, respectively. (d) Fermi surface of $\mathrm{KV_2Se_2O}$, with blue and red curves denoting spin-down and spin-up sectors.}
\end{figure}

We now apply the symmetry and transport framework developed above to $\mathrm{KV_2Se_2O}$ with C-type magnetic configuration. As discussed in Sec.~\ref{sec:symmetry}, the material crystallizes in the tetragonal space group $P4/mmm$ and belongs to the $d$-wave altermagnetic class, with an altermagnetic ordering temperature reported to exceed room temperature \cite{Jiang2025}. As shown in Fig.~\ref{fig:structure_bands}, the magnetic moments are primarily localized on the V atoms and adopt a collinear antiferromagnetic arrangement, with opposite spin orientations on the two sublattices aligned along the $c$ axis. The C-type stacking features no $c$-axis translation that would map the opposite-spin sectors onto each other, and the nonmagnetic O atoms similarly break the effective in-plane translation connecting the two spin sublattices. The remaining spin-space operation $[\mathcal{K}\Vert C_{4z}]$ relates the two opposite-spin sectors and generates the $d$-wave altermagnetic spin splitting shown in Fig.~\ref{fig:structure_bands}(c).

The altermagnetic symmetry manifests clearly in the spin-resolved band structure. Near the X and Y points, the bands display a spin splitting of approximately $1.6\,\mathrm{eV}$, while symmetry enforces spin degeneracies along the $\Gamma$–M direction and, more generally, along the lines $|k_x|=|k_y|$. In the absence of SOC, these degeneracies form spin-group-protected nodal lines characteristic of the altermagnetic phase. The corresponding spin-resolved Fermi-surface crossings along the $|k_x|=|k_y|$ directions are shown in Fig.~\ref{fig:structure_bands}(d). Together, these symmetry and band-structure features make bulk $\mathrm{KV_2Se_2O}$ an ideal material platform for investigating the BCQ-driven nonlinear anomalous Hall effect from first principles.


\subsection{\label{subsec:bcq_distribution}Momentum-space distribution of the BCQ}

\begin{figure}[t]
\centering
\includegraphics[width=\linewidth]{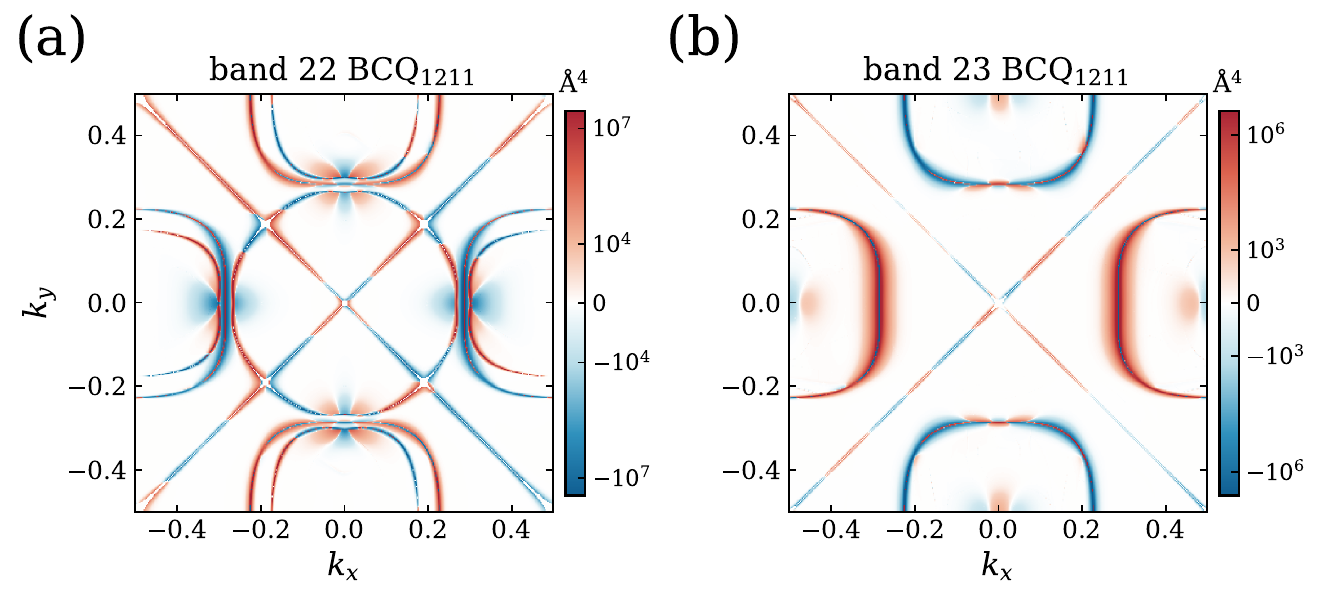}
\caption{\label{fig:bcq_distribution}
Band-resolved momentum-space distribution of the BCQ density
$\Omega^{xy;xx}_n(\mathbf{k})$, in $\text{\AA}^{4}$. Panels show bands $n=22$, $n=23$ in the Wannier-interpolated band
ordering. The calculation is performed on the $k_z=0$ plane.}
\end{figure}

Fig.~\ref{fig:bcq_distribution} shows the band-resolved momentum-space
distribution of the second derivative
$\Omega^{xy;xx}_n(\mathbf{k})$ on the plane $k_z=0$. Panels (a) and (b) correspond to the two bands forming the SOC-induced avoided crossing on this plane (labeled \(n=22\) and \(n=23\) in our band ordering). The two bands exhibit related but not identical hot-spot patterns, consistent with the opposite-sign Berry-curvature contributions expected from an avoided crossing between a pair of bands.

In both panels, sizable BCQ weight tracks the diagonal nodal-line directions $|k_x|=|k_y|$, confirming that the near-degenerate nodal-line region supplies an extended geometric background. Superimposed on this background, sharper arc-shaped features emerge along the $k_x$ and $k_y$ axes, with inner boundaries beginning near $|k_x|\simeq0.2$ (or $|k_y|\simeq0.2$) and extending toward the X and Y regions.

\begin{figure*}[t]
    \centering
    \includegraphics[width=\textwidth]{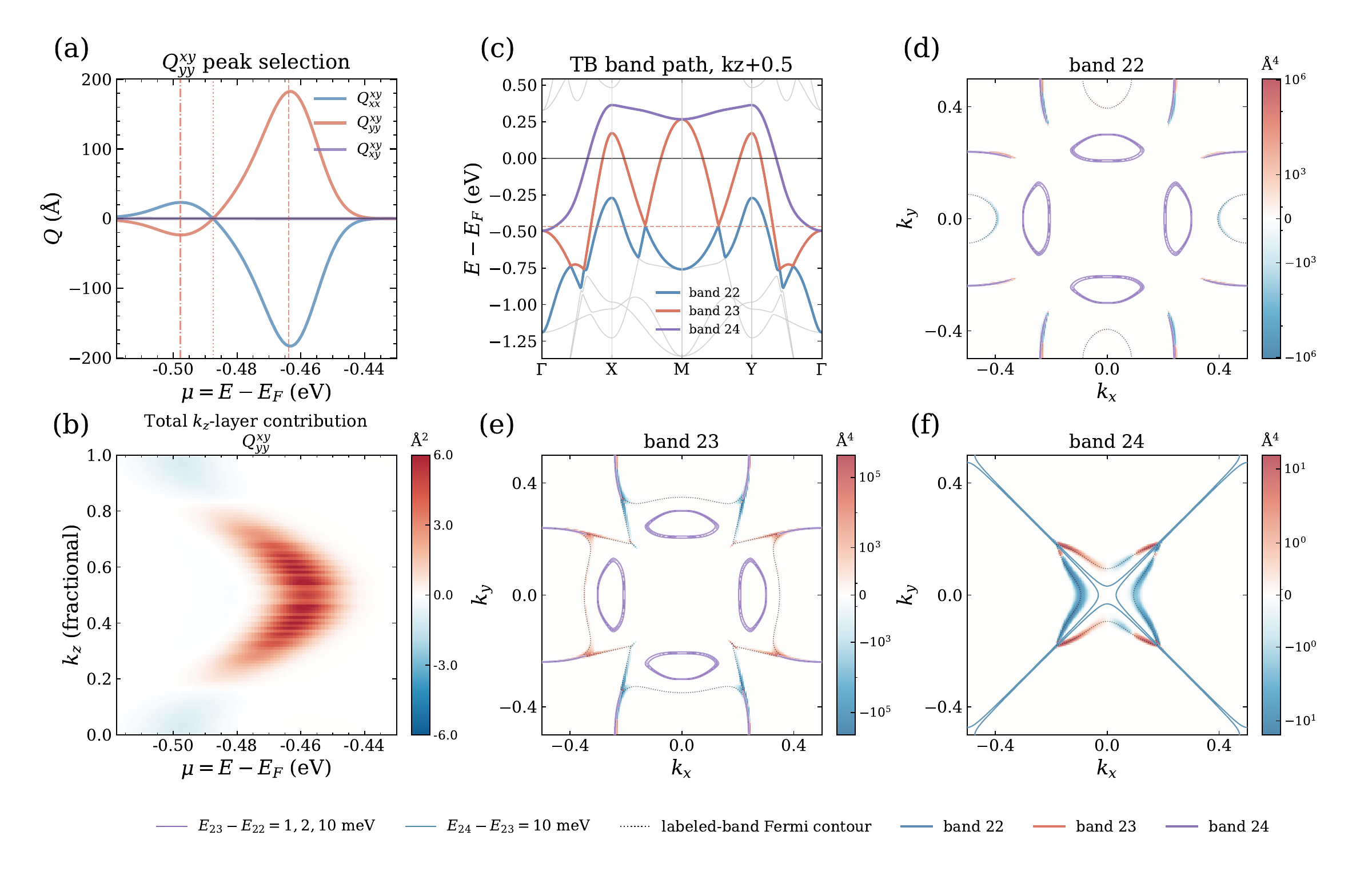}
    \caption{(a) Chemical-potential dependence of the BCQ-induced conductivity coefficient. 
The vertical dashed/dotted lines mark the selected peak, zero-crossing, and valley positions used for the momentum-space decomposition at zero temperature. (b) $k_{z}$ layer-resolve contribution of $Q^{xy}_{yy}$ component 
(c) Band structure along the high-symmetry path, with the horizontal line indicating the energy corresponding to the selected peak chemical potential for the $Q^{xy}_{yy}$ component. 
(d)--(f) Band-resolved momentum-space decomposition of the Fermi-surface weighted integrand 
$\mathrm{BCD}_{122,n}(\mathbf{k})\,v_{2,n}(\mathbf{k})\,\delta_\epsilon(E_n-\mu)$ 
on the $k_z=0.5$ slice for bands 22, 23, and 24, respectively. 
The colored maps show the signed contribution from each band, while the overlaid contours indicate the nearby small-gap structures and the corresponding band-resolved Fermi contours.
    }
    \label{fig:conductivity}
\end{figure*}

The underlying altermagnetic spin splitting and near-degenerate nodal-line geometry are present already in the nonrelativistic limit. Switching on SOC opens small avoided crossings and modifies the spin mixing and interband Berry connections in their immediate vicinity~\cite{Attias2024IntrinsicAHEAltermagnets, Xiang2026ThirdOrderIAHEAltermagnets, PhysRevLett.133.106701}. These small-gap regions host localized BCQ hot spots that are superimposed on the broader diagonal nodal-line background. Upon summing over bands and integrating over the Brillouin zone, the dominant BCQ peak is traced primarily to this avoided-crossing band pair, as discussed in Sec.~\ref{subsec:experimental_estimate}.

\subsection{\label{subsec:experimental_estimate}Estimated signal scale and experimental considerations}

To further elucidate the physical origin of the BCQ-induced nonlinear conductivity, we examine its evolution as a function of the chemical potential. Within the rigid-band approximation, we define the chemical-potential shift relative to the DFT Fermi level as $\mu \equiv E-E_F^{\mathrm{DFT}}$. Fig~\ref{fig:conductivity}(a) shows the resulting dependence of the BCQ conductivity on $\Delta\mu$. In particular, $Q^{xy}_{yy}$ undergoes a sign reversal from negative to positive as the chemical potential is increased. Consistent with the symmetry relation discussed in Sec.~\ref{sec:symmetry}, the BCQ tensor components satisfy $Q^{xy}_{xx}=-Q^{xy}_{yy}$, whereas $Q^{xy}_{xy}$ is strongly suppressed.

To identify the origin of this sign reversal, we decompose the BCQ conductivity into contributions from fixed-$k_z$ slices of the Brillouin zone. Figure~\ref{fig:conductivity}(b) shows that positive and negative contributions originate from distinct $k_z$ regions, and that their magnitudes depend sensitively on the chemical potential. Increasing $\Delta\mu$ enhances the magnitude of the positive-sign slices and suppresses that of the negative-sign slices, which drives the sign change of the integrated response.

The momentum-space origin of these $k_z$-resolved contributions can be examined more closely from their distributions within representative $k_x$-$k_y$ planes. We focus on $\Delta\mu\simeq -0.46~\mathrm{eV}$, where $Q^{xy}_{yy}$ exhibits a pronounced peak, and select $k_z=0.5$ plane, which contributes substantially to the integrated response. The corresponding band dispersion on this plane is shown in Fig.~\ref{fig:conductivity}(c), highlighting the SOC-induced avoided crossing active at this chemical potential.

Figures~\ref{fig:conductivity}(d)–(f) display the associated momentum-space distributions. The blue dashed contours mark $E_{24}-E_{23}=10~\mathrm{meV}$ and delineate the near-nodal band pair, whereas the purple contours trace the small-gap arcs of the SOC-induced avoided crossing between the two bands near the Fermi level ($n=22$ and $n=23$ in our band ordering). At $\Delta\mu\simeq -0.46~\mathrm{eV}$, the dominant BCQ weight is concentrated around the latter manifold and is carried almost entirely by this band pair. The comparison cleanly separates the extended geometric response associated with the near-nodal region from the more localized contribution of the avoided crossing, directly linking the integrated BCQ conductivity to the momentum-space features identified in Fig.~\ref{fig:bcq_distribution}.

The response near the avoided crossing can be further decomposed into the individual band contributions. With increasing $\Delta\mu$, the lower-band contribution is progressively suppressed while the upper-band contribution becomes dominant. Within the rigid-band picture, this crossover reflects the shift of the Fermi level across different portions of the small-gap manifold: the lower branch of the avoided crossing moves away from the Fermi level, while the upper branch is sampled more strongly in a valley-like region of its dispersion. The band-selective momentum-space distributions, shown in Figs.~\ref{fig:conductivity}(d)–(f), confirm this picture. The redistribution is consistent with the $k_z$-resolved evolution discussed above and reflects the $k_z$ dependence of the avoided-crossing gap. A similarly pronounced $k_z$ dependence of Berry-curvature contributions near SOC-gapped nodal structures has been reported in $\mathrm{Fe_3GeTe_2}$~\cite{Kim2018LargeAHEFe3GeTe2}.

The dominant response occurs for a downward chemical-potential shift of approximately $0.4$--$0.5~\mathrm{eV}$. At $\Delta\mu=-0.46~\mathrm{eV}$, the converged value is $|Q^{xy}_{xx}|=180~\text{\AA}$. For a representative thin slab preserving $4_z^{\prime}$ symmetry, Eq.~(\ref{eq:BCQ_conductivity}) with $\hbar/\tau=0.1~\mathrm{eV}$, $E=1~\mathrm{V}/\mu\mathrm{m}$, and an assumed conducting cross section of $1~\mathrm{mm}\times5~\mathrm{nm}$ yields an estimated transverse current of $|I_y^{(3)}|\simeq2.23~\mathrm{mA}$. The complete SI-unit conversion is provided in the Supplemental Material.

The predicted current is a fully quantitative numerical result of the present first-principles calculation under the specified transport parameters. When compared with experiment, however, the sensitivity of the estimate to the avoided-crossing gap, the Hubbard $U$ parameter, the scattering time $\tau$, and the sample geometry, together with the overall $\tau^2 E^3$ scaling, means that it is best interpreted as an order-of-magnitude benchmark rather than a precise experimental prediction. The opposite-sign conductivity peaks flanking the avoided-crossing region nevertheless constitute a robust qualitative feature, in agreement with previous symmetry and model analyses~\cite{PhysRevLett.133.106701,PhysRevB.107.115142}.

\section{Conclusion}

In summary, we have developed a gauge-covariant Wannier-interpolation formalism for evaluating the BCQ in realistic multiband systems, in which the second covariant momentum derivatives of the Berry curvature are expressed analytically in terms of Wannier-gauge quantities, thereby eliminating any reliance on finite differences in momentum space. Applied to the C-type magnetic $\mathrm{KV_2Se_2O}$, the method yields the full momentum-space distribution of the BCQ tensor and reveals a pronounced chemical-potential dependence of the $\tau^2$ BCQ contribution to the third-order Hall conductivity. Band- and momentum-resolved analyses identify an SOC-induced avoided crossing between two spin-split bands near the Fermi level as the dominant microscopic origin of the sign-changing response, with the extended altermagnetic nodal-line regions providing an additional smoother background.
For representative zero-temperature transport parameters and typical device dimensions, we obtain a nonlinear transverse current of approximately $2.23~\mathrm{mA}$, placing the BCQ-driven third-order Hall response in $\mathrm{KV_2Se_2O}$ within the range of experimental sensitivity. More broadly, the formalism implemented in the openly released WannierNLQG package provides a quantitative first-principles route to higher-order quantum-geometric responses in altermagnets and offers a practical foundation for high-throughput screening of candidate materials with strong BCQ-driven nonlinear Hall responses. Identifying robust quantum-geometric descriptors that connect microscopic band features, such as SOC-induced avoided crossings in spin-split altermagnetic bands, to macroscopic nonlinear Hall performance remains an important direction for future work.

\section{Acknowledgments}

H.W. acknowledges the support from the NSFC under Grants Nos. 12474240, 12522411, and 12304049, as well as the support from Fundamental Research Funds for the Central Universities.

\bibliography{main}

\end{document}